\documentclass[conference]{IEEEtran}
\IEEEoverridecommandlockouts
\usepackage[ruled,vlined]{algorithm2e}
\usepackage{subcaption}
\usepackage{bbm}
\usepackage{adjustbox}
\usepackage{lipsum}
\usepackage{stmaryrd}
\usepackage{bm}
\usepackage{cite}
\usepackage{amsmath,amssymb,amsfonts}
\usepackage{algorithmic}
\usepackage{graphicx}
\usepackage{siunitx}
\usepackage{hyperref}
\usepackage{textcomp}
\usepackage{float}
\usepackage{xcolor}
\def\BibTeX{{\rm B\kern-.05em{\sc i\kern-.025em b}\kern-.08em
    T\kern-.1667em\lower.7ex\hbox{E}\kern-.125emX}}
\begin{document}
\title{\huge Joint Satellite Power Consumption and Handover Optimization for LEO Constellations\\
    \thanks{ This work was supported in part by Fonds de recherche du Québec secteur Nature et technologies (FQRNT) and the Natural Sciences and in part by Engineering Research Council (NSERC) of Canada Alliance grant ALLRP~579869-22 ("Artificial Intelligence Enabled Harmonious Wireless Coexistence for 3D Networks (3D-HARMONY)"). }}

\author{Yassine Afif$^*$$^\ddag$, Mohammed Almekhlafi$^*$, Antoine Lesage-Landry$^*$$^\ddag$ and Gunes Karabulut Kurt$^*$\\
$^*$Department of Electrical Engineering, Polytechnique Montréal \& Poly-Grames Research Centre, QC, Canada
\\
$^\ddag$Mila \& GERAD, Montréal, QC, Canada \\
Emails:~$\{$$\text{yassine.afif, mohammed.al-mekhlafi, antoine.lesage-landry, gunes.kurt}\}$@polymtl.ca  
}

\maketitle
\IEEEpeerreviewmaketitle
\begin{abstract}
In satellite constellation-based communication systems, continuous user coverage requires frequent handoffs due to the dynamic topology induced by the Low Earth Orbit (LEO) satellites.
 Each handoff between a satellite and ground users introduces additional signaling and power consumption, which can become a significant burden as the size of the constellation continues to increase. This work focuses on the optimization of the total transmission rate in a LEO-to-user system, by jointly considering the total transmitted power, user-satellite associations, and power consumption, the latter being handled through a penalty on handoff events. We consider a system where LEO satellites serve users located in remote areas with no terrestrial connectivity, and formulate the power allocation problem as a mixed-integer concave linear program (MICP) subject to power and association constraints. Our approach can be solved with off-the-shelf solvers and is benchmarked against a naive baseline where users associate to their closest visible satellite. Extensive Monte Carlo simulations demonstrate the effectiveness of the proposed method in controlling the handoff frequency while maintaining high user throughput. These performance gains highlight the effectiveness of our handover-aware optimization strategy, which ensures that user rates improve significantly, by about 40\%, without incurring a disproportionate rise in the handoff frequency.
\end{abstract}


\section{Introduction}

\IEEEPARstart{T}{he} evolution of wireless communication systems toward 6G is expected to bring unprecedented levels of connectivity, bandwidth, and responsiveness~\cite{giordani2021nonterrestrial}. To enable global coverage and ultra-reliable low-latency communication (URLLC), described in~\cite{feng2021urlcc}, the 3rd Generation Partnership Project (3GPP) has formally included non-terrestrial networks (NTNs) as a key component of its Release 17~\cite{3gpp-38811}. In this context, low Earth orbit (LEO) satellite constellations play a crucial role in providing backhaul and direct access services to mobile and fixed users worldwide~\cite{lin2022path6G}. This technology is particularly valuable in remote, rural, and underserved regions where the deployment of terrestrial infrastructure is impractical or economically unfeasible. As such, LEO satellite technology has been increasingly proposed to address the digital divide and mitigate access inequality~\cite{11021288}.

\begin{figure}[t!]
    \centering
\includegraphics[width=0.46\textwidth]{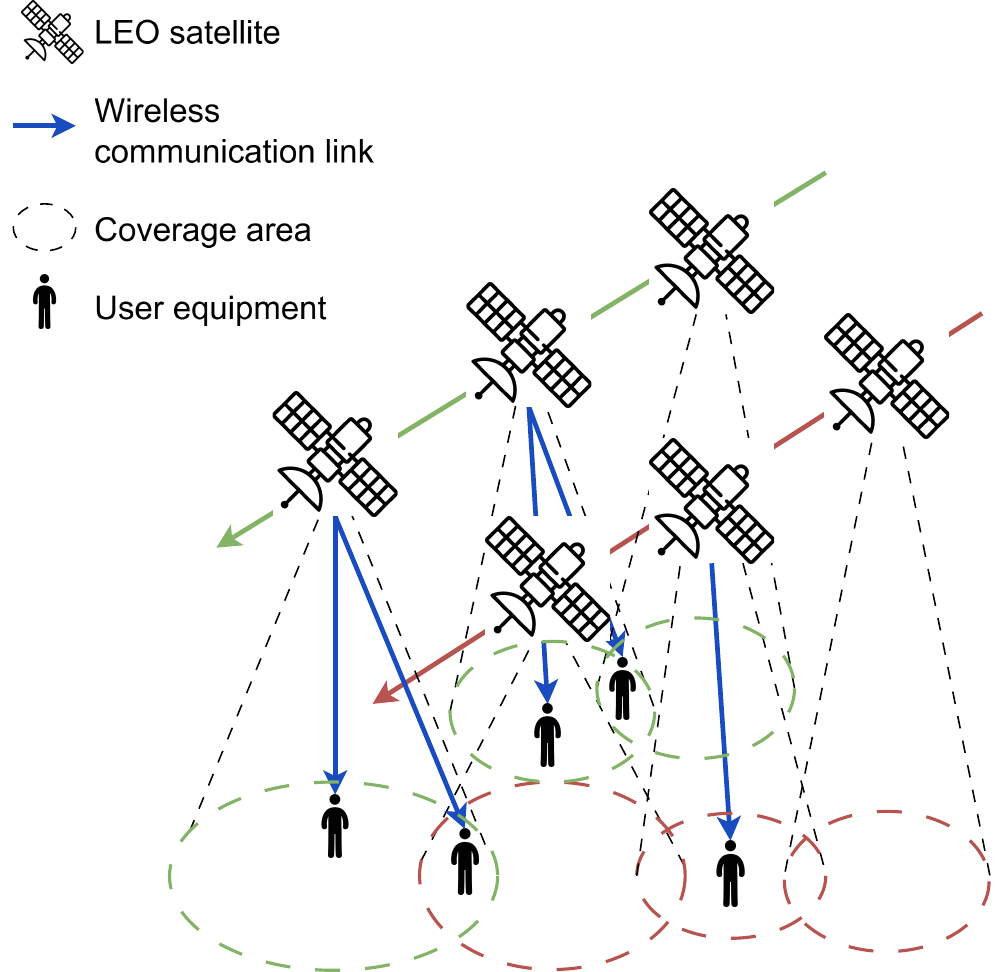}
    \caption{User handovers in a LEO satellite network: moving satellites periodically trigger user reallocation to maintain connectivity.}
    \label{fig:handovergraph}
    \vspace{-.34 in}
\end{figure}

Due to their low orbital altitude, ranging typically around $450$--$2000$ km, LEO satellites exhibit high relative velocities with respect to ground users~\cite{kang2024leo_velocity}. As a result, any user equipment (UE) must undergo frequent handover events, switching between satellite beams or even between satellites to maintain service continuity~\cite{hozayen2022graph_handover}. The term user equipment refers to any device that connects to the wireless network, such as mobile phones, satellite terminals, or Internet of Things (IoT) sensors deployed in the field. Each handover introduces signaling overhead, increased latency, and additional power consumption, the latter of which is especially critical in resource-constrained systems. Therefore, minimizing unnecessary handoffs is essential to preserve system efficiency and user quality of service. A summary of key existing approaches in the literature is provided in Table~\ref{tab:comparison}, highlighting the differences in handover strategies, optimization formulations, and modelling assumptions compared to our work.

\begin{table*}[t]
\centering
\caption{Related Work in LEO Handover Optimization}
\renewcommand{\arraystretch}{1.4}  
\begin{tabular}
{|p{3cm}|p{2.5cm}|p{2.5cm}|p{2.5cm}|p{2.5cm}|}
\hline
\textbf{Aspect} & \cite{voicu2024handover} & \cite{juan2022handover} & \cite{nguyen2023leo} & \textbf{This Work} \\
\hline
Handoff strategy & Closest / visibility / CINR rules & Antenna gain-based handover & Uplink assignment heuristics & Joint optimization of power and handover \\
\hline
Optimization method & Simulation-based evaluation & Antenna gain heuristics & Heuristic resource allocation & Slot-by-slot mixed-integer convex \\
\hline
Channel model & Path loss + visibility zone & Path loss + antenna pattern & Path loss + uplink constraints & Rician fading + path loss + elevation angle \\
\hline
Satellite positions & Simulated geometry (MATLAB) & 3GPP NTN model & Random mobility in 3D space & Real-time Skyfield TLE data \cite{rhodes2019skyfield} \\
\hline
\end{tabular}
\label{tab:comparison}
\end{table*}
This is particularly challenging in dense constellations, where the handoff rate can reach up to multiple events per minute for each user in worst-case scenarios~\cite{alhourani2021session, westphal2023leo_survey,mdpi2025_resource_survey}. A metric used to quantify this behavior is the effective frequency of change (EFC), which measures how often user-satellite associations are updated over time. A high EFC indicates increased signaling overhead and instability in user connections, further motivating the need for mobility-aware resource allocation strategies.

To address the aforementioned challenges, various handover and resource management strategies have been proposed in the literature~\cite{ voicu2024handover, juan2022handover, nguyen2023leo, sun2024handover, hedjazi2012handover,li2020massive}. Reference~\cite{voicu2024handover} analyzed handover strategies in emerging LEO, medium Earth orbit (MEO), and highly elliptical Orbit (HEO) satellite networks, evaluating closest-satellite, maximum visibility, and carrier to interference plus noise ratio (CINR)-based approaches. Their study highlighted key trade-offs between handover rates, spectral efficiency, and propagation delays across different constellation types. Howerver, their work does not consider power consumption or beam-level constraints. The authors of~\cite{juan2022handover} propose antenna gain-based handover heuristics for 5G LEO networks, aiming to reduce control signaling overhead in highly mobile NTN environments. In~\cite{nguyen2023leo}, the authors studied LEO-to-user assignment and resource allocation with the goal of minimizing uplink power. While their approach incorporated realistic channel models and power constraints, handover dynamics and the impact of frequent beam switching on system performance is not considered. Similarly,~\cite{sun2024handover} applies multi-objective reinforcement learning (RL) to optimize handovers in multi-beam LEO networks, achieving adaptive policies that balance user throughput and mobility. While~\cite{voicu2024handover} focuses on optimizing user-satellite association and beamforming to improve handover performance and sum rate in ultra-dense LEO networks, their model does not explicitly account for users in remote areas lacking terrestrial connectivity, nor does it optimize uplink and downlink transmit power jointly during handover events.

Other works address handover strategies from different perspectives. Reference~\cite{hedjazi2012handover} focuses on minimizing handovers in LEO constellations through the optimization of user-satellite association, though they relied on simplified geometrical models. 
Authors in~\cite{li2020massive} investigate massive multiple input multiple output (MIMO) techniques in LEO systems, showing the potential of advanced beamforming to enhance user coverage and mitigate frequent handovers.

While these studies provide valuable insights into handover strategies and resource management, the existing literature falls short of one or more of the following aspects:
\begin{itemize}
    \item They rely on heuristic or rule-based association policies that cannot adapt optimally to dynamic user and satellite configurations.
    \item They focus on small-scale networks or idealized handover models, neglecting the power cost and signaling overheads inherent in large LEO constellations.
    \item They do not jointly consider power allocation and handover minimization in a unified optimization framework.
\end{itemize}

In this paper, we address these gaps by proposing a power-efficient handover optimization framework for medium-scale LEO constellations. Our approach formulates the joint user association and power allocation as a mixed-integer concave program (MICP), i.e., a problem that is concave when binary variables are relaxed to continuous, taking into account beam-level constraints, visibility zones, and handover penalties. We implement a slot-by-slot optimization scheme solvable with off-the-shelf solvers and compare its performance to a naive baseline that assigns each user to the closest visible satellite. Simulation results based on real Starlink data (as of January 2025) illustrates the effectiveness of our method in reducing the handover frequency while maintaining a high user throughput. Our main contributions are summarized as follows:
\begin{itemize}
    \item We develop a scalable simulation framework for user-satellite association using real orbital data and a realistic channel model incorporating Rician fading and propagation losses.
    \item We propose a mixed-integer concave formulation that jointly maximizes total downlink path rate and handover-induced penalties while satisfying the minimum user throughput requirements.
    \item We evaluate the proposed scheme through extensive Monte Carlo simulations and compare it to a baseline strategy in terms of handover frequency, user rate, and overall power efficiency. Our simulations demonstrate an average increase of approximately $40\%$ in per-user downlink throughput compared to the baseline approach, while maintaining a manageable handover rate.
\end{itemize}

The remainder of this paper is organized as follows. Section~\ref{sec:system_model} introduces the system model, formulates the joint downlink optimization problem and outlines the simulation settings with our handover strategy. Section~\ref{sec:results} presents simulation results and performance evaluations. Section~\ref{sec:conclusion} concludes the paper and outlines future research directions.

\section{System Model}
\label{sec:system_model}

In satellite communications, orbiting satellites with high-gain antennas establish links with UE on Earth, enabling connectivity even in remote areas~\cite{maral2011satellite}. User devices communicate within allocated frequency bands, while satellites maintain coverage through multiple beams. In LEOs, the rapid motion of satellites combined with the Earth’s curvature leads to frequent changes in visibility between users and satellites, making continuous service challenging. Adaptative user association and handover strategies are therefore essential to ensure persistent connectivity.

We consider a medium-scale satellite communication system based on a constellation of LEO satellites, each equipped with multiple beams, where each beam represents an independent channel realization rather than a distinct coverage area. The network serves a set of geographically distributed users on Earth over a sequence of discrete time slots \( t = 1, \dots, T,\) where \(T \in \mathbb{N} \) is the time horizon. Due to the dynamic topology of the constellation and the limited visibility of LEO satellites, user-to-beam associations vary frequently, leading to inevitable handovers.

\subsection{System Representation and Handover Modeling}

To formalize the system, let \( \mathcal{S} \subset \mathbb{N}\) denote the set of satellites, \( \mathcal{B} \subset \mathbb{N} \) the set of beams per satellite, which is the same for all satellite, and \( \mathcal{U} \subset \mathbb{N} \) the set of users. Each user can be associated with at most one beam of one satellite at each time slot. This association is captured by a binary variable \( I_{u,s,b}^t \in \{0,1\} \), where \(I_{u,s,b}^t=1\) indicates that user \( u \) is served by beam \( b \) of satellite \( s \) at time~t, and \(I_{u,s,b}^t=0\) otherwise. The visibility of each user is represented by another binary parameter \(V_{u,s,b}^t\), which depends on the relative position of the satellite and the user, as well as the elevation angle. When a satellite is above the horizon, i.e., the elevation angle exceeds a certain threshold, a link can be established, denoted as \(V_{u,s,b}^t=1\); otherwise, \(V_{u,s,b}^t=0\).

User handovers occur when the serving beam or satellite changes between consecutive time slots. These are detected by monitoring changes in \( I_{u,s,b}^t\). Because excessive handovers can degrade the user experience and increase system overhead, we must actively control their frequency. This is because each handover introduces additional signaling overheads, potential service interruptions, and an increased power consumption on both the satellite and user sides, which are particularly limiting in resource-constrained environments such as LEO systems~\cite{hozayen2022graph_handover, alhourani2021session}. Then, a handover penalty \(\alpha > 0\) is introduced in the optimization objective to discourage unnecessary beam switches. This leads to a regularized formulation that balances rate maximization and association changes. The penalty term effectively adds a cost each time a user's association changes from one time slot to the next, thereby promoting more persistent connections. Such regularization is particularly useful in dynamic satellite networks where optimizing only for instantaneous rate may result in excessive and inefficient handovers.

\subsection{Channel and Path Loss Model}

The wireless channel between user \( u \) and beam \( b \) of satellite~\( s \) at time \( t \) is modelled as a Rician fading channel. This model reflects the presence of a strong line-of-sight (LOS) component, characteristic of satellite links, combined with weaker multipath components resulting from atmospheric scattering and reflections. The complex channel coefficient \( h_{u,s,b}^t \) is given by:
\begin{equation}
h_{u,s,b}^t = \sqrt{\frac{K}{K+1}}\, h_{\mathrm{LOS}} + \sqrt{\frac{1}{K+1}}\, h_{\mathrm{NLOS}},
\end{equation}
where \( h_{\mathrm{LOS}} \) represents the deterministic LOS component, and \(h_{\mathrm{NLOS}} \sim \mathcal{CN}(0,1)\) denotes a circularly symmetric complex Gaussian random variable with zero mean and unit variance that captures the random multipath scattering effects. The Rician \( K \)-factor quantifies the relative strength of the LOS and multipath components.

The total path loss \( L_{u,s,b}^t \) incorporates several physical phenomena that attenuate the signal as it propagates through the atmosphere:
\begin{equation}
L_{u,s,b}^t = L_{\mathrm{u,s,b}}^{\text{FS}, t} + L_{\mathrm{u,s,b}}^{\text{atm}, t} + L_{\mathrm{u,s,b}}^{\text{iono}, t} + L_{\mathrm{u,s,b}}^{\text{rain}, t}.
\end{equation}
where:  
\begin{itemize}
\item \( L_{\mathrm{u,s,b}}^{\text{FS}, t} \) is the free-space path loss, which increases logarithmically with the distance between the user and satellite and depends on the carrier frequency. It is computed using the Friis propagation model~\cite{goldsmith2005wireless}:
\begin{equation}
L_{\mathrm{u,s,b}}^{\text{FS}, t} = 32.45 + 20\log_{10}(f_{\text{c}}) + 20\log_{10}(d_{u,s,b}^t),
\end{equation}
With :
\begin{itemize}
    \item \( f_{\text{c}} \) being carrier frequency in MHz,
    \item \( d_{u,s,b}^t \) being distance between user \(u\) and satellite \(s\), beam \(b\) at time slot \(t\), in kilometers.
\end{itemize}
This empirical expression is a widely-used approximation of the Friis transmission formula, 
valid in free-space line-of-sight (LOS) conditions for frequency ranges typically between 
100~MHz and 100~GHz~\cite{goldsmith2005wireless}.

\item  \(L_{\mathrm{u,s,b}}^{\text{atm}, t}\) accounts for the atmospheric absorption, which depends on the elevation angle \(\theta_{u,s,b}^t\) and is modeled as:
\begin{equation}
L_{\mathrm{u,s,b}}^{\text{atm}, t} = \frac{A_{\mathrm{zenith}}}{\sin(\theta_{u,s,b}^t)}.
\end{equation}
with \(A_{\mathrm{zenith}}\) being the the atmospheric attenuation at zenith (i.e., for an elevation angle of \(90^{\circ}\)).

\item \( L_{\mathrm{u,s,b}}^{\text{iono}, t} \) models ionospheric and tropospheric effects. 
Ionospheric effects are significant for frequencies below 6~GHz~\cite{itu618}, 
while tropospheric effects dominate above 6~GHz~\cite{itu618}. 
The corresponding values were interpolated from Table~6.6.6.2.1-1 of~\cite{3gpp-38811}.

\item  \( L_{\mathrm{u,s,b}}^{\text{rain}, t} \) represents rain attenuation. As we consider frequencies below or equal to $20$~GHz, rain attenuation is negligible in a temperate climate 
(such as France), as in our scenario, and we assume rain attenuation is negligible and thus set to zero.
 So we assume \( L_{\mathrm{u,s,b}}^{\text{rain}, t} = 0 \)~\cite{alozie2022rain_att}, for all \(\mathrm{u},\mathrm{s},\mathrm{b}\) and \(\mathrm{t}\).
\end{itemize}
\subsection{SNR and Achievable Rate}

After characterizing the channel and path loss, we compute the signal-to-noise ratio (SNR) for each user as:
\begin{equation}
\gamma_{u,s,b}^t = \frac{|h_{u,s,b}^t|^2}{N_0 W  10^{L_{u,s,b}^t/10}},
\label{eq:SNR}
\end{equation}
where \( N_0 \) is the noise power spectral density and \( W \) is the system bandwidth. The achievable rate for user \( u \) is then given by:
\begin{equation}\label{Rate}
R_{u,s,b}^t = \frac{W}{\log(2)} \log\left(1 + \gamma_{u,s,b}^t P_{u,s,b}^t\right),
\end{equation}
where \( P_{u,s,b}^t \) is the power allocated to user \( u \) at time \( t \).

\subsection{Downlink Optimization Problem}
\label{downlink sec}
In the downlink scenario, the objective is to maximize the total user throughput \(R_u^t\), defined as
\begin{equation}
R_u^t = \sum_{s \in \mathcal{S}} \sum_{b \in \mathcal{B}} R_{u,s,b}^t,
\end{equation}
where \(R_{u,s,b}^t\) is given by~\eqref{Rate} when the link is active and zero otherwise. At each time slot, a handover penalty \(\alpha R_u^{t-1}\) is introduced to discourage frequent user reassignments across beams or satellites, thus balancing instantaneous rate maximization with long-term association stability. The optimization problem can then be formulated as:

\allowdisplaybreaks
{\small
\begin{subequations}\label{downlink}
\begin{align}
\mathcal{P}_t:&\,\max_{\mathbf{P},\mathbf{I}} \sum_{u\in \mathcal{U}} \left(  
   R_{u}^{t} - \alpha  R_{u}^{t-1}  \left(1 - \sum_{s\in\mathcal{S}}\sum_{b\in\mathcal{B}} I_{u,s,b}^{t}  I_{u,s,b}^{t-1} \right) \right), \,\,\nonumber \\ \label{P1:obj} \\
\text{s.t.} 
& \,\, R_{u,s,b}^{t} = \frac{W}{\log(2)} \log\left(1 + \gamma_{u,s,b}^t P_{u,s,b}^t\right), \forall u\in\mathcal{U},\,\nonumber  \\
& \quad s\in\mathcal{S},\, b\in\mathcal{B},\, \label{P1:cons9}
\\
& R_{u}^{t} \geq \gamma_u^{\rm{th}}, \,\, \forall u \in \mathcal{U},\label{P1:cons1} \\
& P_{u,s,b}^{t} \leq I_{u,s,b}^{t} P_s^{\rm{max}}, \,\, \forall u\in\mathcal{U}, s\in\mathcal{S}, b\in\mathcal{B},\label{P1:cons2} \\
& \sum_{u\in\mathcal{U}} P_{u,s,b}^{t} \leq P_s^{\rm{max}}, \,\, \forall s\in\mathcal{S}, b\in\mathcal{B}, \label{P1:cons3}\\
& \sum_{u\in\mathcal{U}} I_{u,s,b}^{t} \leq 1, \,\, \forall s\in\mathcal{S}, b\in\mathcal{B},\label{P1:cons4}\\
& \sum_{s\in\mathcal{S}}\sum_{b\in\mathcal{B}} I_{u,s,b}^{t} \leq 1, \,\, \forall u\in\mathcal{U},\label{P1:cons5}\\
& V_{u,s,b}^{t} \geq I_{u,s,b}^{t}, \,\, \forall u\in\mathcal{U}, s\in\mathcal{S}, b\in\mathcal{B},\label{P1:cons6}\\
& 0 \leq P_{u,s,b}^{t} \leq P_s^{\rm{max}}, \,\, \forall u\in\mathcal{U}, s\in\mathcal{S}, b\in\mathcal{B},\label{P1:cons7}\\
& I_{u,s,b}^{t} \in \{0,1\}, \quad \forall u\in\mathcal{U}, s\in\mathcal{S}, b\in\mathcal{B},\label{P1:cons8}
\end{align}
\end{subequations}
}
The constraints above describe different aspects of the system operation.  Constraint~\eqref{P1:cons9} defines the achievable rate expression \(R_{u,s,b}^t\) as a function of the allocated power and channel gain, based on the Shannon capacity formula. Constraint~\eqref{P1:cons1} guarantees that each user \(u\) achieves a minimum data rate \(\gamma_u^{\rm{th}}\) at every time slot.  
Constraint~\eqref{P1:cons2} enforces that the power allocated to a user on a given beam and satellite does not exceed the maximum power if that beam is selected.
 Constraint~\eqref{P1:cons3} limits the total power transmitted by each satellite on any beam to remain within its maximum capacity, whereas Constraint~\eqref{P1:cons4} ensures that each beam serves at most one user simultaneously. Moreover, Constraint~\eqref{P1:cons5} restricts each user to be associated with at most one beam and satellite per time slot, and Constraint~\eqref{P1:cons6} guarantees that users are only assigned to beams that are visible to them. Constraint~\eqref{P1:cons7} defines the feasible domain of the power variable, ensuring their physical realizability. Finally,~\eqref{P1:cons8} constraint imposes the association variables to binary values.

\begin{figure}[t!]
    \centering
    \begin{subfigure}[b]{0.48\linewidth}
        \centering
        \includegraphics[width=\linewidth]{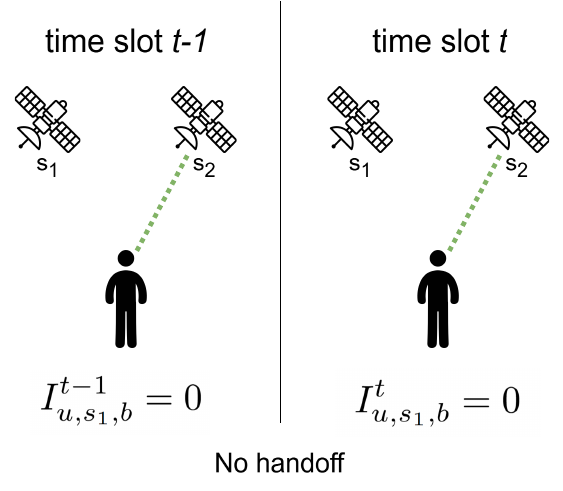}
        \caption{$I_{u,s_{1},b}^{t-1}I_{u,s_{1},b}^t=0$}
        \label{fig:case00}
    \end{subfigure}
    \hfill
    \begin{subfigure}[b]{0.48\linewidth}
        \centering
        \includegraphics[width=\linewidth]{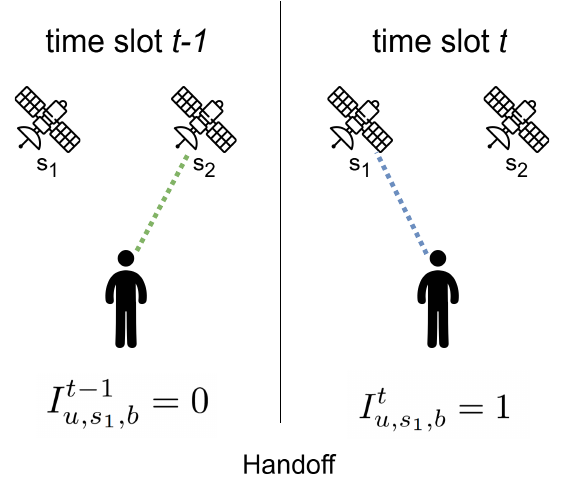}
        \caption{$I_{u,s_{1},b}^{t-1}I_{u,s_{1},b}^t=0$}
        \label{fig:case01}
    \end{subfigure}

    \vspace{0.4cm}

    \begin{subfigure}[b]{0.48\linewidth}
        \centering
        \includegraphics[width=\linewidth]{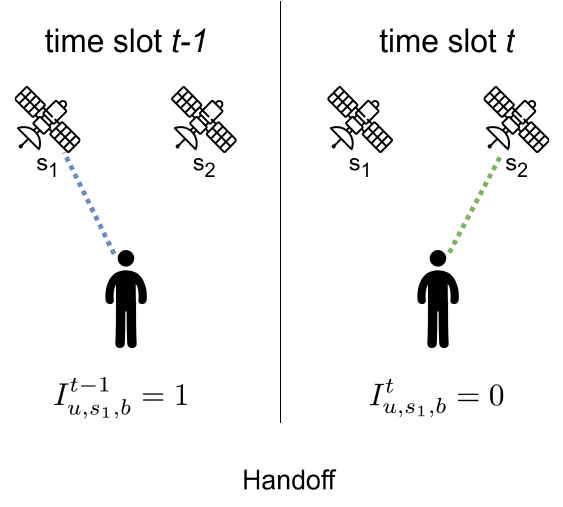}
        \caption{$I_{u,s_{1},b}^{t-1}I_{u,s_{1},b}^t=0$}
        \label{fig:case10}
    \end{subfigure}
    \hfill
    \begin{subfigure}[b]{0.48\linewidth}
        \centering
        \includegraphics[width=\linewidth]{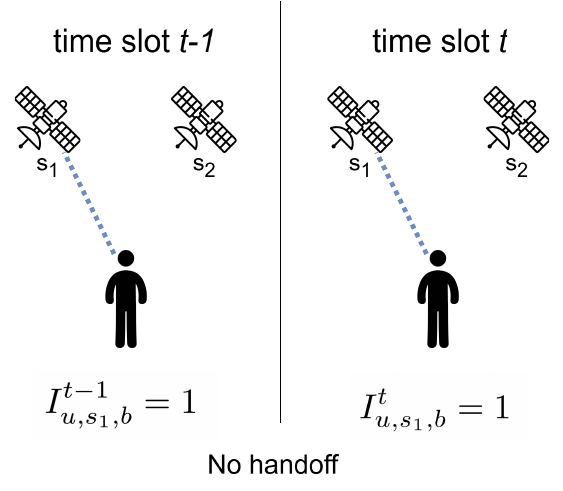}
        \caption{$I_{u,s_{1},b}^{t-1}I_{u,s_{1},b}^t=1$}
        \label{fig:case11}
    \end{subfigure}

    \caption{Illustration of the four possible cases for the product $I_{u,s_{1},b}^{t-1} I_{u,s_{1},b}^t$. 
The binary product is used to detect handovers.}
    \label{fig:I_product_cases}
\end{figure}

Next, we discuss the objective function of~\eqref{P1:obj}. We provide a schematic representation of the product of $I_{u,s_{1},b}^t$ and $I_{u,s_{1},b}^{t-1}$ in Fig.~\ref{fig:I_product_cases}, which illustrates the four possible scenarios depending on user-satellite association over two consecutive time slots.
We analyze the four possible cases shown in Fig.~\ref{fig:I_product_cases}. This binary product helps determine whether a handoff occurs between two consecutive time slots. When $I_{u,s_{1},b}^{t-1} I_{u,s_{1},b}^t = 1$ (Fig.~\ref{fig:case11}), it means the user is served by the same satellite-beam pair at both times $t-1$ and $t$. Consequently, $1 - I_{u,s_{1},b}^{t-1} I_{u,s_{1},b}^t = 0$, and no handoff penalty is applied, as expected. Then, if $I_{u,s_{1},b}^{t-1} I_{u,s_{1},b}^t = 0$ , then $1 - I_{u,s_{1},b}^{t-1} I_{u,s_{1},b}^t = 1$, and a potential handoff penalty may be triggered. However, among the three remaining cases where the product is zero, only two correspond to actual handoff events:
(i) when $I_{u,s_{1},b}^{t-1} = 1$ and $I_{u,s_{1},b}^t = 0$ (Fig.~\ref{fig:case10}, disconnection) and
(ii) when $I_{u,s_{1},b}^{t-1} = 0$ and $I_{u,s_{1},b}^t = 1$ (Fig.~\ref{fig:case01}, new connection). In the fourth case where both $I_{u,s_{1},b}^{t-1} = 0$ and $I_{u,s_{1},b}^t = 0$ (Fig.~\ref{fig:case00}), no service is provided at either time slot, meaning the user is and was not covered at all. Because $I_{u,s_{2},b}^{t-1}$ = $I_{u,s_{2},b}^{t}$ = 1, we will have the same scenario as Fig.~\ref{fig:case11}, and so there will be no handoff there.\\
The double summation \(\sum_{s\in\mathcal{S}}\sum_{b\in\mathcal{B}} I_{u,s,b}^{t} I_{u,s,b}^{t-1}\) in~\eqref{P1:obj} is used to account for all possible satellite-beam pairs in the system. This ensures that the approach correctly detects whether a user remains associated with the same satellite and beam between consecutive time slots. Because each user can only be connected to a single satellite-beam pair at a given time, the sum evaluates to either 1 (no handover) or 0 (handover), allowing the penalty term to be applied consistently across all users. 

In sum, $\mathcal{P}$ is formulated as MICP, which can be solved using off-the-shelf solvers to determine optimal user associations and power allocations under handover constraints.

\subsection{Handover Strategy}
\label{sec:handover_strategy}

To address the frequent handovers inherent in LEO satellite networks, we propose a dynamic user-satellite association strategy that jointly optimizes power allocation and user assignment while penalizing unnecessary beam switching events. This approach ensures a trade-off between maximizing instantaneous throughput and maintaining association stability over time.
The algorithm for the downlink case determines optimal user associations and power allocations at each time slot based on the system model and Problem~\eqref{downlink}. The process is summarized in Algorithm~\ref{algo:DL}

\begin{algorithm}[t!]
\DontPrintSemicolon
\KwIn{User set $\mathcal{U}$, satellite set $\mathcal{S}$, beam set $\mathcal{B}$, channel coefficients $h_{u,s,b}^t$, visibility mask $V_{u,s,b}^t$, power constraints $P_s^{\max}$}
\KwOut{User-satellite-beam associations $\mathbf{I}^t$, allocated powers $\mathbf{P}^t$, achieved rates $\mathbf{R}^t$, handover statistics}
\textbf{Initialization:} Set $\mathbf{I}^0$ using a minimum distance approach, initialize rates $\mathbf{R}^0=0$, and set handover counters $\mathbf{H_u}=0, \, \forall u \in \mathcal{U}$\;
\For{each time slot $t \in \mathcal{T}$}{
    Compute the SNR values $\gamma_{u,s,b}^t$ using~\eqref{eq:SNR}\;
    
    Solve Problem \eqref{downlink}\;

    Update the association matrix $\mathbf{I}^t$ and compute user rates $\mathbf{R}^t$\;

    Detect handovers by comparing $\mathbf{I}^{t}$ and $\mathbf{I}^{t-1}$ as described in Section~\ref{downlink sec}\;
}
\KwRet{$\{\mathbf{I}^t, \mathbf{P}^t, \mathbf{R}^t, \mathbf{H_u}\}$}
\caption{Handover-aware Dynamic User-Satellite Association (Downlink)}
\label{algo:DL}
\end{algorithm}

\vspace{0.3em}

\section{Numerical Results}
\label{sec:results}

\begin{table}[t!]
\centering
\caption{Simulation Parameters}
\label{tab:parameters}
\begin{tabular}{p{4.5cm}p{3.8cm}} 
\hline
\textbf{Parameter} & \textbf{Value} \\
\hline
Satellite TLE data &  (positions, elevation) \\
Start date & 2025/01/01, 00:00 \\
User locations & La Creuse (France), $10$ km radius \\
 Users & $|\mathcal{U}| = 30$ \\
Satellites & $|\mathcal{S}| = 30$ \\
Beams per satellite & $|\mathcal{B}| = 3$ \\
Satellite power & $1$ kW total \\
Visibility angle & $20^\circ$ \\
Carrier frequency & $20$ GHz \\
Bandwidth & $200$ MHz \\
Noise density $N_0$ & $10^{-20}$ W/Hz \\
Rician $K$ factor & $10$ dB \\
Handoff penalty $\alpha$ & $0.5$ \\
Time slots & $20$ \\
Min rate $\gamma^{\rm th}$ & $0.1$ Mbps/user \\
Solver & \texttt{MOSEK} \cite{mosek} \\
\hline
\end{tabular}
\end{table}
In this section, we present the simulation results to evaluate the performance of the proposed handover-aware power allocation algorithm in comparison with a baseline minimum distance association policy, where each user is assigned to the closest visible satellite and beam without considering handover costs or power optimization.
 All results are obtained using simulation parameters provided in Table~\ref{tab:parameters} \cite{m_pimrc}.

\subsection{Throughput Comparison Over Time Slots}
\begin{figure}[b]
\centering
\includegraphics[width=0.48\textwidth]{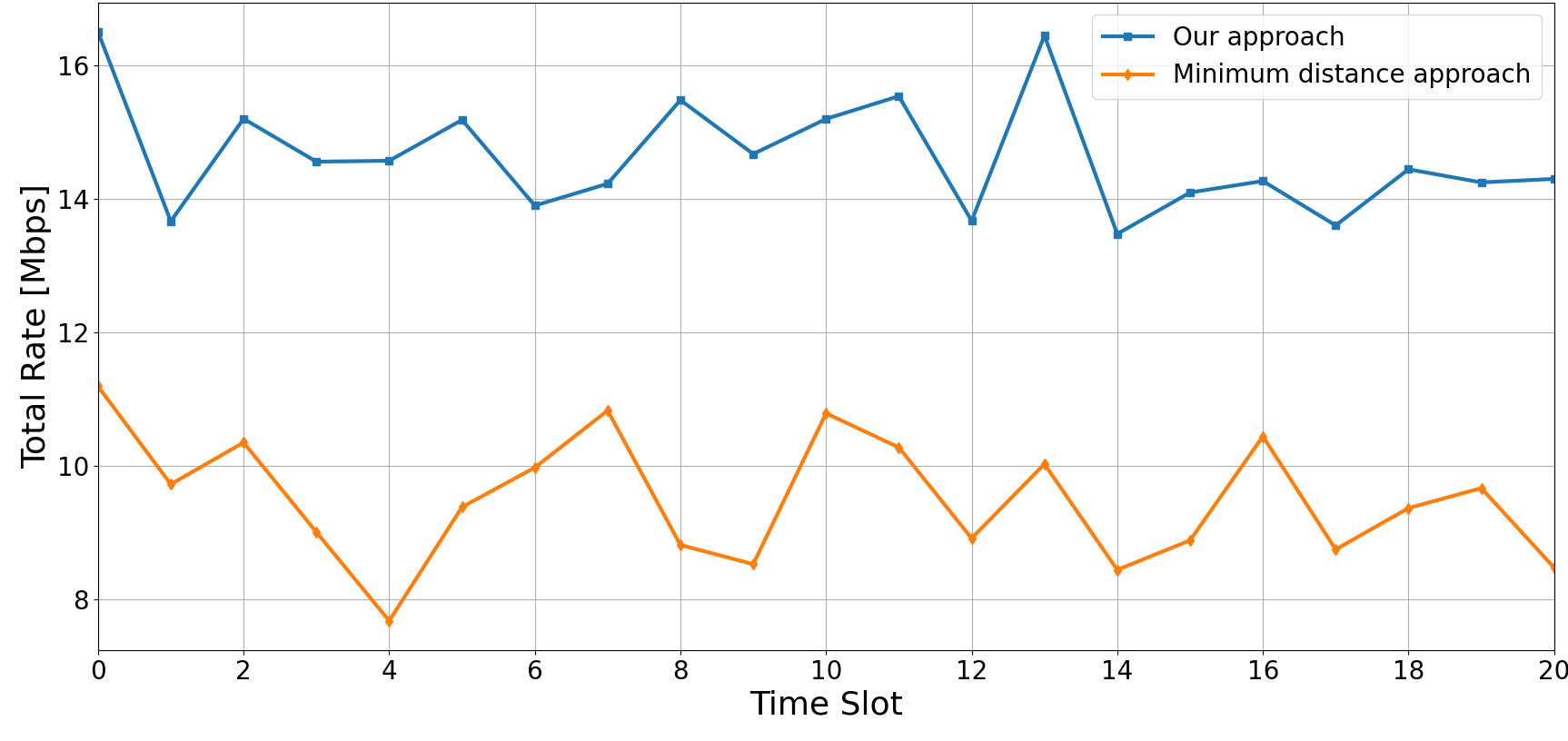}
\caption{Total throughput of all users over $20$ time slots for both approaches.}
\label{fig:throughput_time}
\end{figure}
Fig.~\ref{fig:throughput_time} shows the evolution of the total system throughput (sum of all users' rates) over $20$ time slots for both the proposed approach (blue) and the baseline minimum distance policy (orange). As illustrated, our framework achieves a higher total rate across all users at each time slot, demonstrating its ability to effectively allocate power and associate users with satellites in a way that maximizes spectral efficiency.

\subsection{Monte Carlo Simulation Statistics}
\begin{figure}[t!]
\centering
\includegraphics[width=0.48\textwidth]{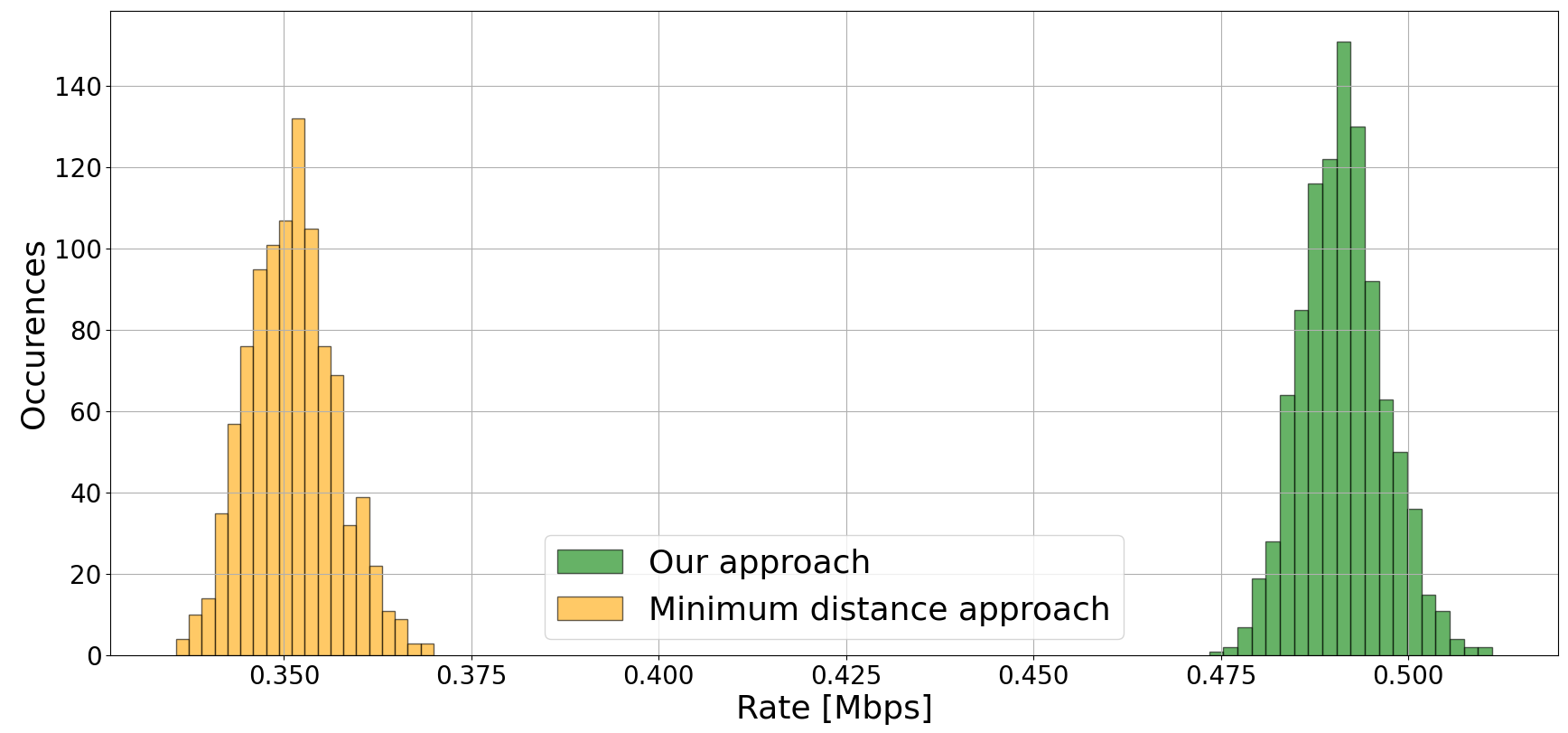}
\caption{Monte Carlo simulation results over $1000$ runs: user rates for both approaches.}
\label{fig:montecarlo_stats}
\end{figure}

We observe an increase in the total rate from about $9.5$ Mbps under the naive baseline to $15$ Mbps with our method, i.e., a gain of nearly $58$\%. This illustrates the benefit of jointly optimizing user association and power allocation instead of simply connecting to the closest satellite. Since this result is from a single run, we perform $1000$ Monte Carlo realizations to assess robustness and account for channel randomness. Over \(T=10\) slots, the baseline achieves an average of $0.351$ Mbps per user, while our approach reaches $0.491$ Mbps, corresponding to a $40$\% improvement (Table~\ref{tab:montecarlo_stats}).

\begin{table}[H]
\centering
\caption{Monte Carlo Simulation Statistics ($1000$ runs)}
\label{tab:montecarlo_stats}
\begin{tabular}{p{3.5cm}p{1.5cm}p{2.5cm}} 
\hline
\textbf{Metric} & \textbf{Proposed Approach} & \textbf{Minimum Distance Approach} \\
\hline
Mean user rate [Mbps] & $0.491$ & $0.351$ \\
Standard deviation [Mbps] & $0.062$ & $0.113$ \\
\hline
\end{tabular}
\end{table}

Fig.~\ref{fig:montecarlo_stats} summarizes these results. The proposed approach achieves a higher average user rate with lower standard deviation, indicating more consistent service quality across users.

\subsection{Handover Analysis}
\begin{figure}[!t]
\centering
\includegraphics[width=0.48\textwidth]{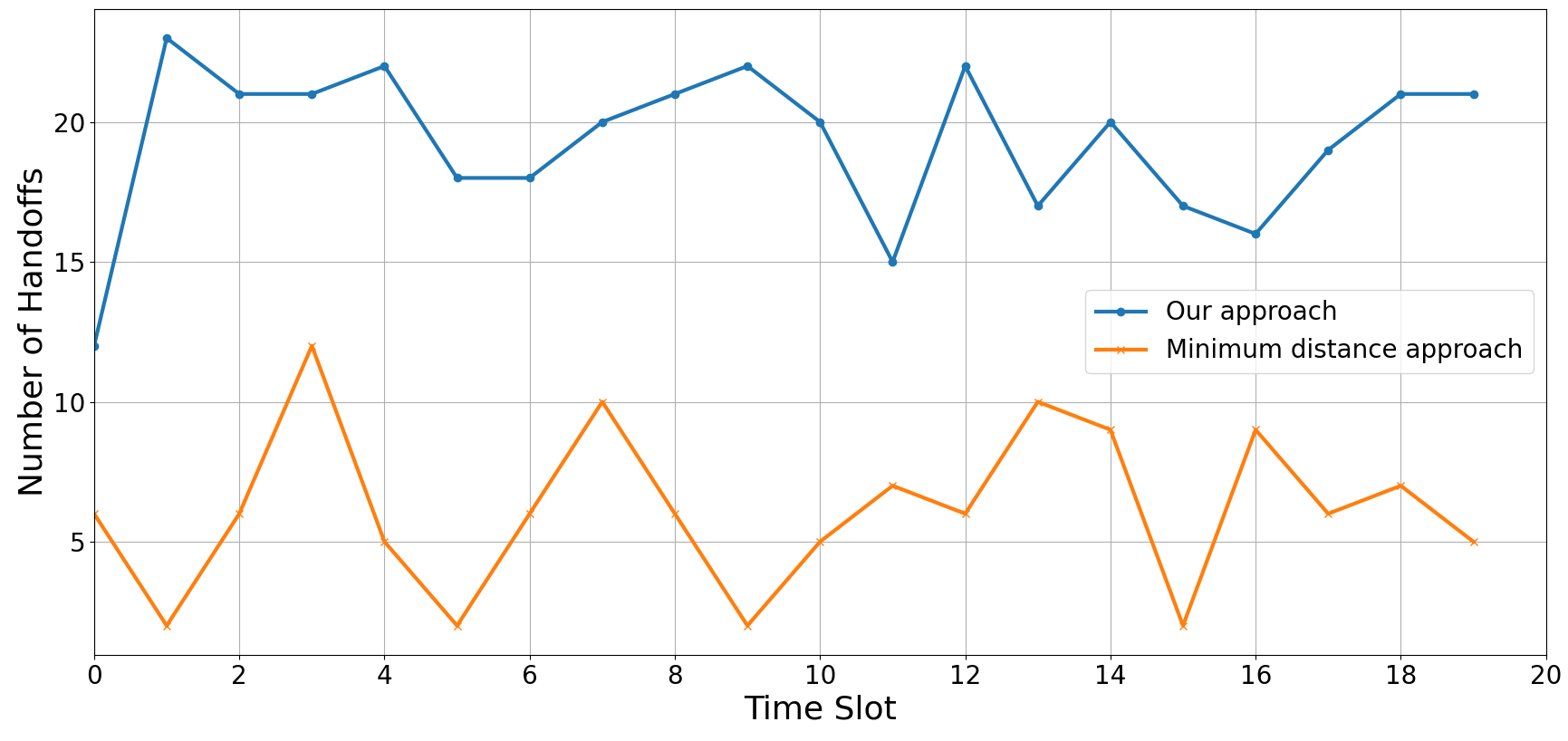}
\caption{Comparison of total number of handovers for the proposed approach and the minimum distance policy.}
\label{fig:handover_comparison}
\vspace{-.3 in}
\end{figure}

While the proposed method outperforms the baseline in terms of throughput, it implies a significantly higher number of handovers, as shown in Fig.~\ref{fig:handover_comparison}. This is due to the optimization algorithm frequently reassigning users to different satellites or beams to maximize the instantaneous rate.

\section{Conclusion}
\label{sec:conclusion}

Our approach showcases the advantages of slot-by-slot joint optimization for maximizing user throughput in Low Earth Orbit (LEO) satellite networks.
While this fine-grained optimization effectively maximizes user throughput, it leads to a significantly higher number of handovers. This is because users are reassigned at each time slot to improve instantaneous rates, without considering longer-term consistency. A promising direction for future work is to perform optimization over more coarse-grained time intervals or over the entire time horizon to reduce handovers, although its non-convexity and sequential nature pose major challenges. Alternatively, decomposition methods such as block coordinate descent could be explored to jointly minimize satellite power consumption and handover frequency in large-scale constellations and very-large networks.

\bibliographystyle{IEEEtran}
\bibliography{refs}

@techreport{3gpp-38811,
  title        = {{3GPP TR 38.811 v15.4.0: Study on New Radio (NR) to support non-terrestrial networks}},
  institution  = {3GPP},
  year         = {2020},
  note         = {Technical Report},
  url          = {https://www.3gpp.org/DynaReport/38811.htm}
}

@inproceedings{nguyen2023leo,
  title     = {{{LEO}-to-User Assignment and Resource Allocation for Uplink Transmit Power Minimization}},
  author    = {Nguyen-Kha, Hung and Ha, Vu Nguyen and Lagunas, Eva and Chatzinotas, Symeon and Grotz, Joel},
  booktitle = {Proc. IEEE WSA \& SCC},
  year      = {2023},
}

@article{sun2024handover,
  title={{Handover for Multi-Beam {LEO} Satellite Networks: A Multi-Objective Reinforcement Learning Method}},
  author={Sun, Yang and Zhai, Yuqing and Wu, Wenjun and Si, Pengbo and Yu, Fei Richard},
  journal={IEEE Commun. Lett.},
  volume={28},
  number={12},
  year={2024},
  pages     = {2834--2838}
}

@inproceedings{hedjazi2012handover,
  title     = {{Optimization of the Problem of Handover in Low Earth Orbit Satellite Constellations}},
  author    = {Hedjazi, {Nour el Houda} and Ouacifi, Malika and Bouchouareb, Rachida and Ouarghi, Meriem and Benatia, Djamel and Gareh, Messaoud},
  booktitle = {Proc. IEEE SETIT},
  year      = {2012},
}

@INPROCEEDINGS{li2020massive,
  author    = {You, Li and Li, Ke-Xin and Wang, Jiaheng and Gao, Xiqi and Xia, Xiang-Gen and Ottersten, Bj{\"o}rn},
  title     = {{LEO} Satellite Communications With Massive {MIMO}},
  booktitle = {2020 IEEE International Conference on Communications (ICC)},
  year      = {2020},
  doi       = {10.1109/ICC40277.2020.9149121},
  publisher = {IEEE},
  address   = {Dublin, Ireland},
  month     = jun
}

@ARTICLE{voicu2024handover,
  author={Zhou, He and Li, Jianguo and Yang, Kai and Zhou, Haibo and An, Jianping and Han, Zhu},
  journal={IEEE Trans.  Veh. Technol.}, 
  title={{Handover Analysis in Ultra-Dense LEO Satellite Networks With Beamforming Methods}}, 
  year={Mar. 2023},
  volume={72},
  number={3},
  pages={3676-3690},
}

@article{juan2022handover,
  author={Juan, Enric and Lauridsen, Mads and Wigard, Jeroen and Mogensen, Preben},
  journal={IEEE Access}, 
  title={{Handover Solutions for 5G Low-Earth Orbit Satellite Networks}}, 
  year={2022},
  volume={10},
  number={},
  pages={93309-93325}
}

@book{goldsmith2005wireless,
  title     = {{Wireless Communications}},
  author    = {Goldsmith, Andrea},
  publisher = {Cambridge University Press},
  year      = {2005},
  address   = {Cambridge, UK},
  isbn      = {9780521837163},
  doi       = {10.1017/CBO9780511841224}
}

@techreport{itu618,
  title        = {{Propagation Data and Prediction Methods Required for the Design of Earth–Space Telecommunication Systems}},
  institution  = {{International Telecommunication Union (ITU)}},
  type         = {{ITU-R Recommendation}},
  number       = {P.618-14},
  year         = {2023},
  url          = {https://www.itu.int/rec/R-REC-P.618},
}

@article{giordani2021nonterrestrial,
  title     = {{Non-Terrestrial Networks in the 6{G} Era: Challenges and Opportunities}},
  author    = {Marco Giordani and Michele Zorzi},
  journal   = {IEEE Network},
  year      = {Mar. 2021},
  volume    = {35},
  number    = {2},
  pages     = {244--251},
  doi       = {10.1109/MNET.011.2000405},
  publisher = {IEEE}
}

@article{feng2021urlcc,
  title     = {{Ultra-Reliable and Low-Latency Communications: Applications, Opportunities and Challenges}},
  author    = {Feng, Daquan and Lai, Lifeng and Luo, Jingjing and Zhong, Yi and Zheng, Canjian and Ying, Kai},
  journal   = {Science China Information Sciences},
  year      = {Jan. 2021},
  volume    = {64},
  article-number = {120301},
  doi       = {10.1007/s11432-020-2852-1},
  publisher = {Science China Press}
}

@article{lin2022path6G,
 author={Lin, Xingqin and Cioni, Stefano and Charbit, Gilles and Chuberre, Nicolas and Hellsten, Sven and Boutillon, Jean-Francois},
  journal={IEEE Commun. Mag.}, 
  title={{On the Path to 6G: Embracing the Next Wave of Low Earth Orbit Satellite Access}}, 
  year={Dec. 2021},
  volume={59},
  number={12},
  pages={36-42},
}

@article{kang2024leo_velocity,
  title     = {{Downlink Analysis of a Low-{E}arth Orbit Satellite Considering an Airborne Interference Source Moving on Various Trajectory}},
  author    = {Kang, Eunjung and Park, YoungJu and Kim, JungHoon and Choo, Hosung},
  journal   = {Remote Sensing},
  year      = {Jan. 2024},
  volume    = {16},
  number    = {2},
  pages     = {321},
  doi       = {10.3390/rs16020321},
  publisher = {MDPI}
}

@inproceedings{hozayen2022graph_handover,
  title        = {{A Graph-Based Customizable Handover Framework for {LEO} Satellite Networks}},
  author       = {Hozayen, Mohamed and Darwish, Tasneem and Karabulut Kurt, Gunes and Yanikomeroglu, Halim},
  booktitle    = {Proc. IEEE GLOBECOM },
  year         = {2022},
  month        = nov,
  pages        = {868--873},
  doi          = {10.1109/GCWkshps56602.2022.10008514},
}

@article{alhourani2021session,
  title     = {{Session Duration between Handovers in Dense Low {E}arth Orbit Satellite Networks}},
  author    = {Al-Hourani, Akram},
  journal   = {IEEE Commun. Lett.},
  year      = {2021},
  month     = {oct},
  volume    = {10},
  number    = {12},
  pages     = {2810--2814},
  doi       = {10.1109/LWC.2021.3118214},
  publisher = {IEEE}
}

@article{westphal2023leo_survey,
  title     = {{{LEO} Satellite Networking Relaunched: Survey and Current Research Challenges}},
  author    = {Westphal, Cedric and Han, Lin and Li, Richard},
  journal   = {ITU Journal on Future and Evolving Technologies},
  year      = {2023},
  month      = {oct},
  volume    = {4},
  number    = {4},
  doi       = {10.52953/LWXC1928},
}

@article{mdpi2025_resource_survey,
  title        = {{Hierarchical Resource Management for Mega-{LEO} Satellite Constellation: A Review}},
  author       = {Gou, Liang and Bian, Dongming and Nie, Yulei and Zhang, Gengxin and Zhou, Hongwei and Shi, Yulin and Zhang, Lei},
  journal      = {Sensors},
  year         = {2025},
  volume       = {25},
  number       = {3},
  article-number = {902},
  publisher    = {MDPI}
}

@article{alozie2022rain_att,
  title     = {{A Review on Rain Signal Attenuation Modeling, Analysis and Validation Techniques: Advances, Challenges and Future Direction}},
  author = {Alozie, Emmanuel and Abdulkarim, Abubakar and Abdullahi, Ibrahim and others},
  journal   = {Sustainability},
  year      = {2022},
  volume    = {14},
  number    = {18},
  doi       = {10.3390/su141811744},
  publisher = {MDPI}
}

@book{maral2011satellite,
  title={{Satellite Communications Systems: Systems, Techniques and Technology}},
  author={Maral, Gerard and Bousquet, Michel},
  publisher={John Wiley \& Sons},
  edition={5},
  year={2011}
}

@ARTICLE{11021288,
  author={Almekhlafi, Mohammed and Lesage-Landry, Antoine and Karabulut Kurt, Gunes },
  journal={IEEE Open J. Veh. Technol.}, 
  title={{Access Inequality in {LEO} Satellite Networks: A Case Study of High-Latitude Coverage in Northern Québec}}, 
  year={May 2025},
  volume={6},
  number={},
  pages={1613-1630},
  doi={10.1109/OJVT.2025.3575546}}

@INPROCEEDINGS{m_pimrc,
  author={Almekhlafi, Mohammed and Lesage-Landry, Antoine and Karabulut Kurt, Gunes }, 
    booktitle={ Proc. IEEE PIMRC}, 
  title={{Connectivity-Aware Task Offloading for Remote Northern Regions: a Hybrid {LEO}-{MEO} Architecture}}, 
  year={Sep.  2025},
  address = { Istanbul, Türkiye}}

@manual{mosek,
   author = "MOSEK ApS",
   title = {{The MOSEK optimization toolbox for MATLAB manual. Version 9.0.}},
   year = 2019,
   url = "http://docs.mosek.com/9.0/toolbox/index.html"
 }

@article{rhodes2019skyfield,
  title={{Skyfield: High precision research-grade positions for planets and Earth satellites generator}},
  author={Rhodes, Brandon},
  journal={Astrophysics Source Code Library},
  pages={ascl--1907.024},
  year={2019}
}

\end{document}